# Soliton excitation in waveguide arrays with an effective intermediate dimensionality

A. Szameit[1], Y. V. Kartashov[2], F. Dreisow[1], M. Heinrich[1], T. Pertsch[1], S. Nolte[1], A. Tünnermann[1], V. A. Vysloukh[3], F. Lederer[4], and L. Torner[2]

[1]*Institute of Applied Physics, Friedrich Schiller University Jena, Max-Wien-Platz 1, 07743 Jena, Germany*

[2]*ICFO-Institut de Ciencies Fotoniques, and Universitat Politecnica de Catalunya, Mediterranean Technology Park, 08860 Castelldefels (Barcelona), Spain*

[3]*Departamento de Fisica y Matematicas, Universidad de las Americas – Puebla, 72820, Puebla, Mexico*

[4]*Institute for Condensed Matter Theory and Optics, Friedrich-Schiller-University Jena, Max-Wien-Platz 1, 07743 Jena, Germany*

We reveal and observe experimentally significant modifications undertaken by discrete solitons in waveguide lattices upon the continuous transformation of the lattice structure from one-dimensional to two-dimensional. Light evolution and soliton excitation in arrays with a gradually increasing number of rows are investigated, yielding solitons with an effective reduced dimensionality residing at the edge and in the bulk of the lattice.

A continuously renewed interest is devoted to physical systems with built-in reduced dimensionality, such as quantum wells, quantum wires and quantum dots [1] or periodic and quasi-periodic semiconductor superlattices [2], where the artificial confinement of the electronic wave function yields a variety of specific phenomena [3]. Such confinement is usually studied in condense-matter physics, where the band structure is a result of the periodic atomic potential. However, due to the mathematical analogy describing the evolution of electronic and photonic wave packages, many related phenomena can be explored better in the optical domain. Waveguide arrays are an exceptional model system for quantum mechanical processes (see e.g. [4,5]). In this Letter we demonstrate in an optical lattice how confinement of the electromagnetic wave function in one direction results in a modification of the light evolution in the linear as well as in the nonlinear regime. We show that in such a system intermediate states can exist that do not occur in pure one-dimensional (1D) or in two-dimensional (2D) geometries. Similar effects may appear in magnetic nonlinear chains, molecular crystals, nonlinear metamaterials, or matter-waves held in optical lattices with variable effective dimensionality.

In an optical periodic medium, the evolution of stationary nonlinear excitations is fundamentally different from the evolution in continuous media. The interplay between nonlinearity, diffraction, and periodic modulation of refractive index may result in the formation of so-called discrete solitons [6], which have been observed in a variety of materials, such as semiconductors [7], optically induced lattices in photorefractive materials [8], and waveguide arrays permanently written in fused silica with femtosecond laser pulses [9]. The theoretical investigations included not only the simplest fundamental solitons, but also discrete vortices [10], soliton trains [11], necklaces [12], and spatio-temporal bullets [13]. One salient outcome of such investigations is that the properties of solitons in periodic media depend critically on the lattice geometry and symmetry [14]. On the other hand, a fundamental property of all types of solitons is that their features greatly depend on dimensionality. For example, even in homogeneous Kerr (cubic) media, one-dimensional fundamental solitons exist for any non-vanishing power levels and are all stable, while higher-dimensional soliton solutions exist only above a power threshold and are all unstable.

Here we address discrete surface solitons residing near the edge and at the edge of periodic lattices. Such states *exist only above a threshold power* both, for one-dimensional [15] and for two-dimensional [16,17] lattices. The power threshold depends on the distance between the surface and the excitation [18], on the form-anisotropy of the coupling along different directions in the lattice [16], and on the geometry of the lattice interface itself [17]. While the formation of fundamental discrete solitons in pure 1D and 2D geometries in Kerr-type media is a well explored topic, the modification of the soliton properties, shapes, symmetries, as well as their power thresholds for existence, have never been studied for a gradual transition from 1D to 2D waveguiding geometries, neither theoretically nor experimentally. Higher-dimensional discrete solitons are known to require higher powers for their exci-tation, but the dependence and behavior of the corresponding thresholds on an intermediate dimensionality and the conditions at which the arrays resemble truly 1D or 2D settings, are unknown. In this Letter we show why and how the thresholds for the existence of fundamental discrete solitons in Kerr media change upon modification of the effective dimensionality of the guiding structure, and at which conditions the structure can be considered as effectively 1D or 2D. We study solitons in the center of the array, as well as edge and corner surface excitations.

To study the impact of the lattice confinement on the light propagation we considered a sequence of arrays with a gradually increasing number of rows in the vertical direction. Each row contains 7 channels. This allows one to consider all intermediate situations between effectively 1D ($7 \times 1$ array) and 2D systems ($7 \times 7$ array), characterized by the variable ratio of number of rows in horizontal and vertical directions (for example $7 \times 2$ array). We describe the propagation of a light beams along the $\xi$-axis in these systems with the nonlinear Schrödinger equation for the dimensionless light field amplitude $q$ under an assumption of cw radiation:

$$i\frac{\partial q}{\partial \xi} = -\frac{1}{2}\left(\frac{\partial^2 q}{\partial \eta^2} + \frac{\partial^2}{\partial \zeta^2}\right) - |q|^2 q - pR(\eta,\zeta)q. \tag{1}$$

Here $\eta, \zeta$ and $\xi$ are normalized transverse and longitudinal coordinates; the parameter $p$ describes the refractive index modulation; the function $R(\eta,\zeta)$ characterizes the lattice profile that is approximated by two Gaussian functions $\exp[-(\eta/w_\eta)^2 - (\zeta/w_\zeta)^2]$ fitting the profiles of two waveguides separated by a distance $w_s$. In agreement with $4.5 \times 12$ $\mu m^2$ cross-sections of our laser-written waveguides we set $w_\eta = 4.5$ and $w_\zeta = 1.2$. The separation between waveguides in our samples is $40$ $\mu m$ (i.e., $w_s = 4.0$). The nonlinearity strength is $n_2 = 2.7 \times 10^{-20}$ $m^2/W$. The refractive index modulation depth is $p = 2.9$ which corresponds to a refractive index contrast of $\sim 3.2 \times 10^{-4}$.

First of all, we studied the impact of the lattice dimensionality on the properties of stationary soliton solutions of the form $q(\eta,\zeta,\xi) = w(\eta,\zeta)\exp(ib\xi)$, where $b$ is the propagation constant that is connected to the total power $U = \int_{-\infty}^{\infty} |q|^2 \, d\eta d\zeta$ of a soliton. Representative profiles of solitons residing in the bulk, corner and edge of the array are shown in Fig. 1. All such solitons exist above a cutoff $b_{co}$ that increases with increasing index modulation $p$ and also with increasing number of rows. Close to the cutoff, the soliton width diverges, which is in particular the case in the 1D $7 \times 1$ array where both bulk and surface solitons strongly expand across the array [Figs. 1(a) and 1(c)]. The increasing peak amplitude that accompanies a growth of $b$ causes a progressive localization in a single site [Fig. 1(b)]. When the number of rows in the array is slightly increased, the light can expand also along the vertical direction. One can obtain surface and bulk solitons that occupy many sites in horizontal direction, but are confined to only several sites (given by the number of rows) in vertical direction [Figs. 1(d)-1(i)]. Such solitons realize an intermediate situation between purely 1D or 2D states. The larger the number of rows in the array the stronger the soliton expansion close to the cutoff, so that in the $7 \times 7$ array one gets a true 2D solitons on a square lattice [Fig. 1(g)-1(i)]. Importantly, the degree of soliton expansion in the vertical direction impacts dramatically the corresponding $U(b)$ dependence (Fig. 2). While in a one-dimensional $7 \times 1$ system the power of the central soliton residing in the array bulk is a monotonically increasing function of $b$ [Fig. 2(a), curve 1], in a system with more than one row there exist regions where the slope $dU/db$ of $U(b)$ curve becomes negative, so that local minima form [Fig. 2(a), curves 2 and 3]. Since we are dealing with a spatially limited array that supports extended linear modes, the soliton can not expand indefinitely and as a result the power eventually vanishes in the cutoff. Still, the $U$ value in the local minimum almost exactly coincides with the power threshold for solitons in the corresponding horizontally infinite array. One can see that this local threshold appearing for bulk solitons already in the $7 \times 2$ array increases with the number of rows upon transition into a 2D system. Surface states exhibit power thresholds even in 1D [18], and we found that the threshold increases rapidly with the transition from 1D to 2D [Fig. 2(b)]. Thus, in our arrays for corner solitons the threshold power increases monotonically from $U_{\text{th}} = 0.236$ for $7 \times 1$ array to $U_{\text{th}} = 0.496$ in $7 \times 7$ array. Although the threshold for bulk and surface excitation strictly increase with growing number of layers, the rates of increase are different [Fig. 2(c)]. While in 1D system surface solitons exhibit a higher threshold than bulk solitons, at approximately 3 layers the thresholds are similar, so that the curves intersect. When the number of layers is further increased, the threshold for surface states approaches asymptotically the limit $U_{\text{th}} = 0.496$ already at 4 layers, while the threshold for bulk solitons further increases and reaches higher values. Hence, the situation is reversed for weak and strong vertical confinement.

It should be stressed that the appearance of surface modes like those depicted in Figs. 1(c), 1(e), 1(f) and Fig. 3(c)-3(f) is directly related to the phenomenon of symmetry breaking that is possible at sufficiently high powers. Thus, besides symmetric and antisymmetric modes concentrated in the bulk of the guiding structure it can support an asymmetric mode that may be localized near the surface, but only if the power exceeds certain minimum value. Such symmetry breaking is possible in simple layered structures as it was predicted in [19]. The asymmetric modes are known to branch out of symmetric or antisymmetric modes and are characterized by a nonmonotonic $U(b)$ dependence [see Fig. 2(b) and Fig. 1 of Ref. [19]].

To explore our results experimentally, we fabricated a sequence of arrays consisting of 1, 2, ..., 7 rows using the femtosecond direct-writing technique [20]. The writing velocity was $\sim 1750$ $\mu$m/s to ensure the spatially uniform nonlinearity [21]. In a first step we compared the degree of localization at intermediate input power at 1.0 MW for a different number of layers (Fig. 3). When exciting a waveguide in the bulk, in the pure 1D case (one layer), the localization is already well pronounced [Fig. 3(a)]. However, adding a second layer allows light to couple also into the vertical direction [Fig. 3(c)]. As a consequence, the light penetrates strongly into the second row, resulting in the formation of a symmetric mode of the two-layer system (a linear mode of this type is also supported by this structure) and in an overall decrease of localization, while in one-layer system the light was already well localized at this power level [Fig. 3(a)]. The picture is further changed with three layers. When exciting the central waveguide, the degree of localization in the horizontal direction is reduced, while the light couples into the adjacent rows [Fig. 3(e)]. In the 2D system (seven layers), light spreads over the entire system in the horizontal and vertical directions and almost no localization is obtained [Fig. 3(g)]. We observed a different behavior when a corner waveguide is excited. In the 1D system localization is again well pronounced [Fig. 3(b)]. In the two-layer system a slight tendency to delocalization in horizontal direction and the formation of a vertical symmetric mode can be observed [Fig. 3(d)]. The influence of the finiteness in the vertical direction is still visible for the three-layer-system. Importantly, in this case the localization in the horizontal direction is significantly reduced compared to the one- and two-layer systems [Fig. 3(f)]. Finally, in 2D cases (seven layers), the excited modes are strongly extended and feature long tails penetrating into the depth of array [Fig. 3(h)].

To gain insight into the formation of solitons in arrays with varying dimensionality, the excitation of the corner waveguide in the lowest row of the three-layer system was investigated. The outcome is shown in Fig. 4. For the corresponding simulations we solved Eq. (1) with input $q = A\exp[-(\eta^2 + \zeta^2)/W^2]$, where $A$ and $W$ are the input beam amplitude and width, respectively. In the linear regime, at 0.16 MW, light linearly couples into the highest row and scatters off the surface [Fig. 4(a)]. For an intermediate power of 0.7 MW, the repulsive force of the surface is almost overcome and light is attracted by the surface waveguide [Fig. 4(b)]. For high enough input power (1.5 MW), a well localized soliton forms [Fig. 4(c)]. Importantly, note that the behavior of light for a corner excitation is different in a true 2D system. Namely: At low-powers (0.16 MW) light is strongly repelled from both surfaces [Fig. 5(a)]. Since there is no confinement in horizontal as well as in vertical direction, light gradually contracts towards the excited waveguide at intermediate power with the typical exponential tails [Fig. 5(b)]. At high input power (1.7 MW), a surface soliton is generated [Fig. 5(c)]. A comparison between Figs. 4 and 5 makes readily visible the increase of the threshold for soliton formation.

In conclusion, we addressed soliton formation in a system characterized by a substantially different transverse extent of light along the horizontal and the vertical axes. We observed solitons corresponding to the transition between purely 1D and 2D systems. We demonstrated the different soliton shapes and threshold powers required for the excitation of discrete bulk and surface solitons upon the transition from 1D to 2D structures.

# Figure captions

Figure 1. Profiles of central solitons with $b = 0.562$ (a) and $b = 0.740$ (b), and corner soliton with $b = 0.567$ (c) in $7 \times 1$ array. Profiles of central soliton with $b = 0.604$ (d), edge soliton with $b = 0.605$ (e), and corner soliton with $b = 0.609$ (f) in $7 \times 3$ array. Profiles of central soliton with $b = 0.612$ (g), edge soliton with $b = 0.613$ (h), and corner soliton with $b = 0.612$ (i) in $7 \times 7$ array.

Figure 2. Power versus propagation constant for (a) solitons residing in central waveguide of $7 \times 1$ (curve 1), $7 \times 3$ (curve 2), $7 \times 7$ (curve 3) arrays, and (b) solitons residing in the corner waveguide of $7 \times 1$ (curve 1), $7 \times 2$ (curve 2), $7 \times 7$ (curve 3) arrays. Points marked by circles in (a) correspond to solitons shown in Figs. 1(a), 1(g), and 1(b), while points in (b) correspond to solitons in Figs. 1(c) and 1(i). (c) Threshold power for solitons residing in corner (white circles) and central (black circles) waveguides versus number of layers.

Figure 3. Partial light localization at 1.0 MW in a central waveguide [panels (a), (c), (e), and (g)] and in a corner waveguide [panels (b), (d), (f), and (h)]. Panels (a) and (b) correspond to $7 \times 1$ array, panels (c) and (d) to $7 \times 2$ array, panels (e) and (f) to $7 \times 3$ array, and panels (g) and (h) correspond to $7 \times 7$ array.

Figure 4. Excitation of a corner soliton in $7 \times 3$ array. (a) 0.16 MW, (b) 0.7 MW and (c) 1.5 MW. Top row: experiment. Bottom row: theory.

Figure 5. Excitation of a corner soliton in $7 \times 7$ array. (a) 0.16 MW, (b) 0.7 MW and (c) 1.7 MW. Top row: experiment. Bottom row: theory.

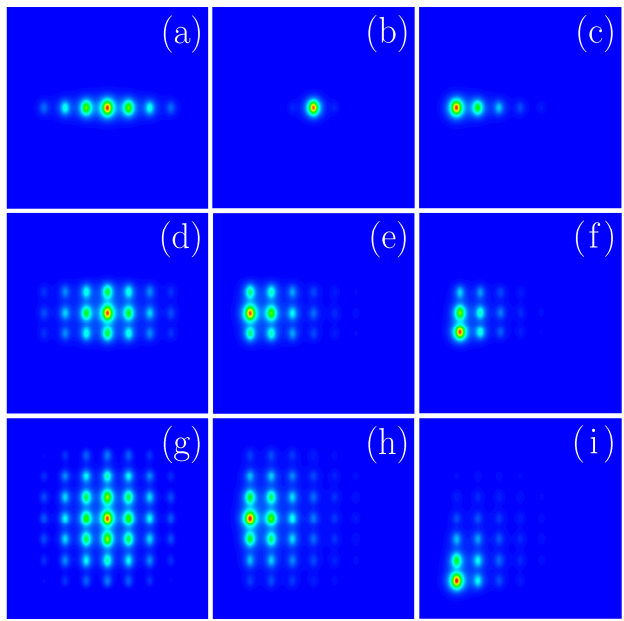

Figure 1. Profiles of central solitons with $b = 0.562$ (a) and $b = 0.740$ (b), and corner soliton with $b = 0.567$ (c) in $7 \times 1$ array. Profiles of central soliton with $b = 0.604$ (d), edge soliton with $b = 0.605$ (e), and corner soliton with $b = 0.609$ (f) in $7 \times 3$ array. Profiles of central soliton with $b = 0.612$ (g), edge soliton with $b = 0.613$ (h), and corner soliton with $b = 0.612$ (i) in $7 \times 7$ array.

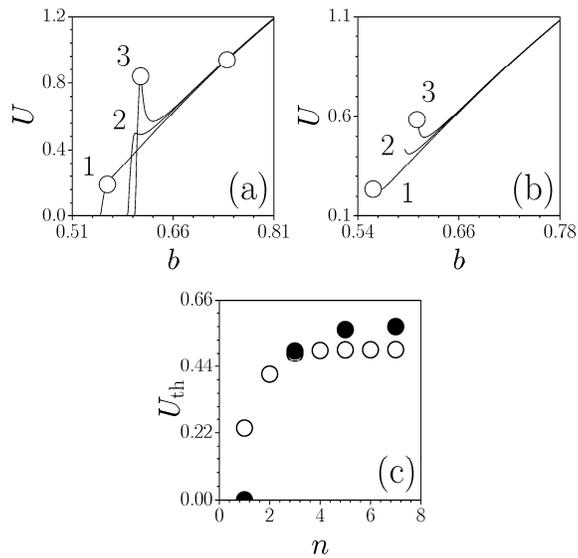

Figure 2. Power versus propagation constant for (a) solitons residing in central waveguide of $7 \times 1$ (curve 1), $7 \times 3$ (curve 2), $7 \times 7$ (curve 3) arrays, and (b) solitons residing in the corner waveguide of $7 \times 1$ (curve 1), $7 \times 2$ (curve 2), $7 \times 7$ (curve 3) arrays. Points marked by circles in (a) correspond to solitons shown in Figs. 1(a), 1(g), and 1(b), while points in (b) correspond to solitons in Figs. 1(c) and 1(i). (c) Threshold power for solitons residing in corner (white circles) and central (black circles) waveguides versus number of layers.

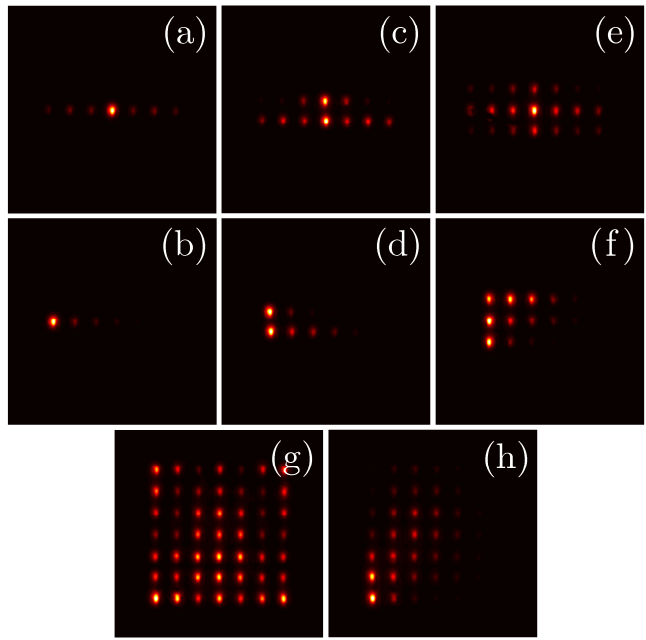

Figure 3. Partial light localization at 1.0 MW in a central waveguide [panels (a), (c), (e), and (g)] and at 0.7 MW in a corner waveguide [panels (b), (d), (f), and (h)]. Panels (a) and (b) correspond to $7 \times 1$ array, panels (c) and (d) to $7 \times 2$ array, panels (e) and (f) to $7 \times 3$ array, and panels (g) and (h) correspond to $7 \times 7$ array.

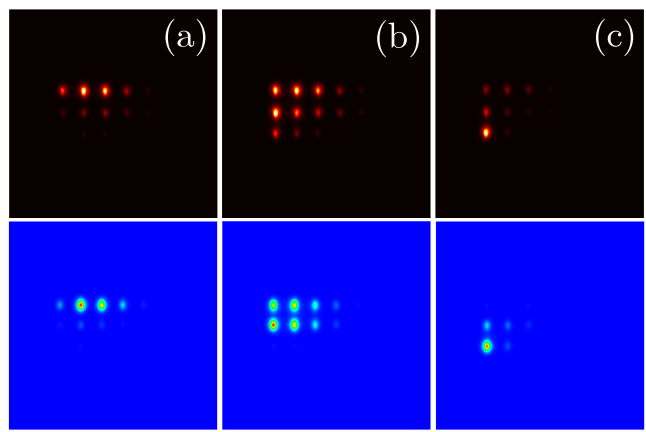

Figure 4. Dynamical excitation of a corner soliton in $7 \times 3$ array. (a) 0.16 MW, (b) 0.7 MW and (c) 1.5 MW. Top row: experiment. Bottom row: theory.

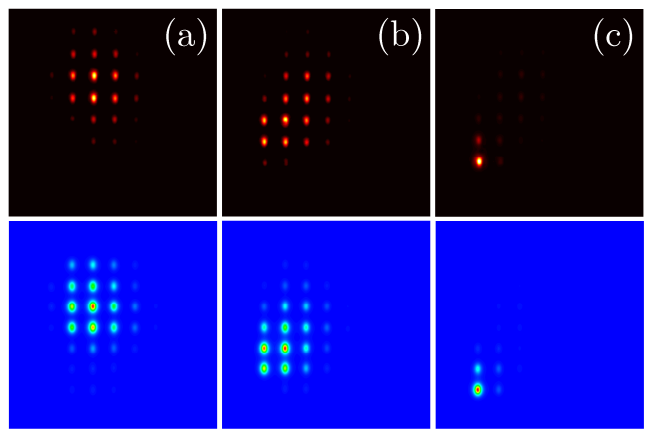

Figure 5.  Dynamical excitation of a corner soliton in $7 \times 7$ array. (a) 0.16 MW, (b) 0.7 MW and (c) 1.7 MW. Top row: experiment. Bottom row: theory.